\documentclass[11pt]{article}

\usepackage{color}
\usepackage{latexsym}
\usepackage{amssymb}
\usepackage{graphicx}
\usepackage{amsmath, amsfonts,amssymb,theorem,euscript,array,enumerate,amsfonts,mathrsfs}

\newtheorem{Theorem}{Theorem}[part]

\newtheorem{Proposition}{Proposition}[part]

\newtheorem{Remark}{Remark}[part]

\def\esssup_#1{\underset{#1}{\mathrm{ess\,sup\, }}}
\def\essinf_#1{\underset{#1}{\mathrm{ess\,inf\, }}}
\def\argmax_#1{\underset{#1}{\mathrm{arg\,max\, }}}

\def \Frac{\displaystyle\frac}

\def \R{\mathbb{R}}

\def \E{\mathbb{E}}
\def \F{\mathbb{F}}

\def \P{\mathbb{P}}
\def \Q{\mathbb{Q}}

\def \Ac{{\cal A}}

\def \Ec{{\cal E}}
\def \Fc{{\cal F}}

 \def \Nc{{\cal N}}

\def \eps{\varepsilon}

\def \ep{\hbox{ }\hfill$\Box$}

\def\DT#1{\Frac{\partial #1}{\partial T}}

\def\reff#1{{\rm(\ref{#1})}}

\def\beqs{\begin{eqnarray*}}
\def\enqs{\end{eqnarray*}}
\def\beq{\begin{eqnarray}}
\def\enq{\end{eqnarray}}

\begin{document}

\title{Long time asymptotics  for optimal investment\thanks{Contribution to Springer Proceedings in Asymptotic Methods in Finance (Editors Friz-Gatheral-Gulisashvili-Jacquier-Teichmann), in memory of Peter Laurence. }}

\author{Huy\^en PHAM
             \\\small  Laboratoire de Probabilit\'es et
             \\\small  Mod\`eles Al\'eatoires
             \\\small  CNRS, UMR 7599
             \\\small  Universit\'e Paris Diderot
             \\\small  pham at math.univ-paris-diderot.fr
             \\\small  and CREST-ENSAE 
             }


\maketitle

\vspace{20mm}

\begin{abstract}
This survey reviews portfolio selection problem for long-term horizon. We consider  two  objectives:  (i)  maximize the probability  for outperforming   a target growth rate of wealth process  (ii) minimize the probability of falling below a target growth rate. We  study the asymptotic behavior of these criteria formulated as   large deviations control pro\-blems, that we solve by duality method leading to ergodic risk-sensitive portfolio optimization problems. 
Special emphasis is placed  on linear factor models where explicit solutions  are obtained.  
\end{abstract}

\vspace{5mm}

\noindent {\bf MSC Classification (2000):} 60F10, 91G10, 93E20.

\vspace{5mm}

\noindent {\bf Keywords:}  Long-term investment, large deviations, risk-sensitive control, ergodic HJB equation.

\newpage

\section{Introduction}

 Dynamic portfolio selection looks for strategies maximizing some performance criterion. It is a main topic in mathematical finance, first solved in continuous time in the seminal paper  \cite{mer71}, and extended in various directions by taking into account  stochastic investment opportunities, market imperfections and/or transaction costs.  We refer for instance to the textbooks \cite{kor97}, \cite{karshe98} or \cite{pha09}, and the recent survey paper \cite{liumuh13} for developments on this subject.  

Classical criterion for investment decision is the expected utility maximization from terminal wealth, which requires to specify on one hand the utility function representing the investor's preference, and subjective by nature, and on the other hand the finite horizon.  We consider in this paper an alternative behavioral 
foundation, with an objective criterion over long term. More precisely, we are concerned with the performance of a portfolio relative to a given target, and 
are interested in maximizing (resp. minimizing) the probability to outperform (resp. to fall below)  a target growth rate when time horizon goes to infinity.  
Such criterion, formulated as a large deviations portfolio optimization problem,   
has been proposed by \cite{stu03} in a static framework,  studied in a continuous-time framework for the maximization of upside chance probability 
by \cite{pha03}, and then by \cite{hatsek05}, see also \cite{ste04} in discrete-time models. The asymptotics of minimizing the downside risk probability 
is studied in \cite{hatnagshe10} and \cite{nag12}. 

Large deviations portfolio optimization is a nonstandard stochastic control problem, and is tackled by  duality approach. The dual control problem 
is an ergodic risk-sensitive portfolio optimization problem studied in \cite{fleshe00} by dynamic programming PDE methods in a Markovian setting, see also \cite{guarob12}, and 
leads to particularly tractable results with time-homogenous policies.  
A nice feature of the duality approach is also  to relate the target level in the  objective probability of  upside chance maximization or  downside risk 
minimization to the subjective degree of risk aversion, hence to make endogenous the utility function of the investor.

The rest of this paper is organized as follows. Section 2 formulates  the large deviations criterion. In Section 3, we state the general duality relation 
for the large deviations optimization problem, both for the upside chance probability maximization and downside risk minimization.  
We illustrate in section 4 our results in the Black-Scholes toy model with constant proportion  portfolio. Finally, we consider in Section 5 a factor model for 
assets price, and characterize  the optimal strategy of the large deviations optimization problem via  the resolution of an ergodic Hamilton-Jacobi-Bellman equation from the  risk-sensitive dual control.  Explicit solutions  are provided in the linear Gaussian factor model.

\section{Large deviations criterion}

\setcounter{equation}{0} \setcounter{Assumption}{0}
\setcounter{Theorem}{0} \setcounter{Proposition}{0}
\setcounter{Corollary}{0} \setcounter{Lemma}{0}
\setcounter{Definition}{0} \setcounter{Remark}{0}

We  study  a portfolio choice criterion, which is preferences-free, i.e. objective, and horizon-free, i.e. over long term investment.   
This is formulated as a large deviations criterion that we now describe in an abstract set-up.  
On a filtered probability space $(\Omega,\Fc,\F=(\Fc_t)_{t\geq 0},\P)$ supporting all the random quantities appearing in the sequel, we 
consider a frictionless  financial market with $d$ assets of positive price process $S$ $=$ $(S^1,\ldots,S^d)$. There is  an agent investing at any time $t$ 
a fraction $\pi_t$ of her wealth in the assets based on the available information $\Fc_t$. We denote by $\Ac$ the set of admissible control strategies 
$\pi$ $=$ $(\pi_t)_{t\geq 0}$, and $X^\pi$ the associated positive wealth process of dynamics:
\beq \label{Xpi}
dX_t^\pi &=& X_t^\pi \pi_t'{\rm diag}(S_t)^{-1} dS_t, \;\;\; t \geq 0,
\enq
where ${\rm diag}(S_t)^{-1}$ denotes the diagonal $d\times d$ matrix of $i$-th diagonal term $1/S_t^i$.  We then define the so-called {\it growth rate portfolio}, i.e. the logarithm of the wealth process $X^\pi$: 
\beqs
L_t^\pi & := & \ln X_t^\pi, \;\;\; t \geq 0. 
\enqs
We set  by $\bar L^\pi$ the average growth rate portfolio over time:
\beqs
\bar L_t^\pi &:=& \frac{L_t^\pi}{t}, \;\;\; t > 0. 
\enqs 

We shall then consider two problems on the long time asymptotics for the average growth rate:
\begin{itemize}
\item[(i)] {\bf Upside chance probability}: given a target growth rate $\ell$, the agent wants to maximize over portfolio strategies $\pi$ $\in$ $\Ac$
\beqs
\P\big[ \bar L_T^\pi \geq \ell \big] & & \mbox{ when } \; T \rightarrow \infty. 
\enqs
\item[(ii)] {\bf Downside risk probability}: given a target growth rate $\ell$, the agent wants to minimize over portfolio strategies $\pi$ $\in$ $\Ac$
\beqs
\P\big[ \bar L_T^\pi \leq \ell \big] & & \mbox{ when } \; T \rightarrow \infty. 
\enqs
\end{itemize}
Actually, when horizon time $T$ goes to infinity, the probabilities of upside chance or downside risk have typically an exponential decay in time, and we are led to the following mathematical formulations of large deviations criterion: 
\beq
v_+(\ell) &:=& \sup_{\pi\in\Ac} \limsup_{T\rightarrow\infty} \frac{1}{T} \ln P\big[ \bar L_T^\pi \geq \ell \big],  \label{defv+} \\
v_-(\ell) &:=& \inf_{\pi\in\Ac} \liminf_{T\rightarrow\infty} \frac{1}{T} \ln P\big[ \bar L_T^\pi \leq \ell \big].  \label{defv-}
\enq
This criterion  depends on the objective probability $\P$, and the target growth rate $\ell$, but there is no exogenous utility function, and finite horizon.  
Large deviations control problem \reff{defv+} and \reff{defv-} are nonstandard in the literature on stochastic control, and we shall study these problems by 
a duality approach.

\section{Duality}

\setcounter{equation}{0} \setcounter{Assumption}{0}
\setcounter{Theorem}{0} \setcounter{Proposition}{0}
\setcounter{Corollary}{0} \setcounter{Lemma}{0}
\setcounter{Definition}{0} \setcounter{Remark}{0}

 We derive in this section the dual formulation of the large deviations criterion introduced in \reff{defv+}-\reff{defv-}. 
 Given $\pi$ $\in$ $\Ac$,  if  the average growth rate portfolio $\bar L_T^\pi$ satisfies a large deviations principle, then large deviations theory states 
 that its rate function $I(.,\pi)$ should be related  to its limiting log-Laplace transform $\Gamma(.,\pi)$ by duality via the G\"artner-Ellis theorem:
 \beq \label{dualpi}
 I(\ell,\pi) & = & \sup_\theta \big[ \theta \ell - \Gamma(\theta,\pi) \big],
 \enq
 where $I(.,\pi)$ is the rate function associated to the LDP of $\bar L_T^\pi$:
 \beq \label{rate}
 \limsup_{T\rightarrow\infty} \frac{1}{T} \ln P\big[ \bar L_T^\pi \geq \ell \big] &=& - \inf_{\ell'\geq\ell} I(\ell',\pi) \; = \;  I(\ell,\pi), \;\;\; \ell \geq \lim_{T\rightarrow\infty} \bar L_T^\pi, 
 \enq 
 and $\Gamma(.,\pi)$ is the limiting log-Laplace transform of $\bar L_T^\pi$:
 \beqs
 \Gamma(\theta,\pi) &:=&  \limsup_{T\rightarrow\infty} \frac{1}{T}  \ln \E \big[ e^{\theta T \bar L_T^\pi} \big], \;\;\; \theta \in \R, 
 \enqs
 The issue is now to extend this duality relation \reff{dualpi} when optimizing over control $\pi$.  To fix the ideas, let us formally derive  
 from \reff{dualpi}-\reff{rate} the maximization of upside chance probability. 
 \beqs
 \sup_\pi  \limsup_{T\rightarrow\infty} \frac{1}{T} \ln P\big[ \bar L_T^\pi \geq \ell \big]  &=& \sup_\pi \big[ - I(\ell,\pi) \big] \\
 &=& \sup_\pi \Big[ - \sup_\theta\big[ \theta\ell - \Gamma(\theta,\pi) \big] \Big]  \\
 &=& \sup_\pi \inf_\theta  \big[ \Gamma(\theta,\pi) - \theta\ell \big] \\
\mbox{ (if we can invert  sup and inf)}&=&  \inf_\theta \big[ \sup_\pi \Gamma(\theta,\pi) - \theta\ell \big]. 
 \enqs
 We thus expect that 
 \beq \label{dualv+}
 v_+(\ell) &=&  \inf_\theta \big[ \Lambda_+(\theta) - \theta\ell \big], 
 \enq
 where $\Lambda_+$ is defined by
 \beqs 
 \Lambda_+(\theta) &=& \sup_\pi \Gamma(\theta,\pi). 
 \enqs
 In other words, we should have a duality relation between the value function $v_+$ of the large deviations control problem, and the value function 
 $\Lambda_+$, which is known in the mathematical finance literature, as an ergodic  risk-sensitive portfolio optimization problem.

 Let us now state rigorously the duality relation in an abstract (model-free) setting.  We first consider the upside chance large deviations probability, and define the corresponding dual control problem: 
 \beq \label{defLambda}
 \Lambda_+(\theta) & := & \sup_{\pi\in\Ac}  \limsup_{T\rightarrow\infty} \frac{1}{T}  \ln \E \big[ e^{\theta T \bar L_T^\pi} \big], \;\;\; \theta \geq 0. 
 \enq
 We easily see from H\"older inequality that $\Lambda_+$ is convex on $\R_+$.  The following result is due to \cite{pha03}. 
 
 \begin{Theorem} \label{theodual+}
 Suppose that $\Lambda_+$ is finite and differentiable on $(0,\bar\theta)$ for some $\bar\theta$ $\in$ $(0,\infty]$, and there exists $\hat\pi(\theta)$ $\in$ $\Ac$ solution to 
 $\Lambda_+(\theta)$ for any $\theta$ $\in$ $(0,\bar\theta)$. Then, for all $\ell$ $<$ $\Lambda_+'(\bar\theta)$, we have:
 \beqs
 v_+(\ell) &=& \inf_{\theta \in [0,\bar\theta)} \big[ \Lambda_+(\theta) - \theta\ell \big]. 
 \enqs
Moreover, an optimal control for $v_+(\ell)$, when $\ell$ $\in$ $(\Lambda_+'(0),\Lambda_+'(\bar\theta))$, is
 \beqs
 \pi^{+,\ell} &=&  \hat\pi(\theta(\ell)), \;\;\; \mbox{ with } \; \Lambda_+'(\theta(\ell)) \; = \; \ell,  
 \enqs
 while a nearly-optimal control for $v_+(\ell)$ $=$ $0$, when $\ell$ $\leq$ $\Lambda_+'(0)$, is: 
 \beqs
 \pi^{+(n)} &=& \hat\pi(\theta_n), \;\;\;  \mbox{ with } \;  \theta_n =   \theta\big(\Lambda_+'(0)+ \frac{1}{n}\big) \overset{n\rightarrow\infty}{\longrightarrow} 0,  
 \enqs
 in the sense that
 \beqs
 \lim_{n\rightarrow\infty} \limsup_{T\rightarrow\infty} \frac{1}{T} \ln  \P\big[ \bar L_T^{\pi^{+(n)}} \geq \ell \big] &=& v_+(\ell).
 \enqs

 \end{Theorem}
 {\bf Proof.}  
 \noindent {\it Step 1.}
Let us  consider the Fenchel-Legendre transform of the convex function $\Lambda_+$ on $[0,\bar\theta)$:
\begin{eqnarray} \label{deffenchel}
\Lambda_+^*(\ell) &=&
\sup_{\theta\in [0,\bar\theta)}[\theta \ell  - \Lambda_+(\theta)],
\;\;\; \ell  \in \R.
\end{eqnarray}
Since $\Lambda_+$ is $C^1$ on $(0,\bar\theta)$, it is well-known (see e.g. Lemma 2.3.9 in \cite{demzei98}) that the function  $\Lambda_+^*$ is convex, nondecreasing and satisfies:
\begin{eqnarray} \label{Lam*}
\Lambda_+^*(\ell) &=&
\left\{
\begin{array}{cl}
\theta(\ell) \ell - \Lambda_+(\theta(\ell)), & \mbox{if } \;
 \Lambda_+'(0)  < \ell <  \Lambda_+'(\bar\theta) \\
 0, & \mbox{if } \; \ell  \leq \Lambda_+'(0),
\end{array}
\right.
\end{eqnarray}
\begin{eqnarray} \label{resultexposed}
\theta(\ell) \ell - \Lambda_+^*(\ell) &>& \theta(\ell) \ell'  - \Lambda_+^*(\ell'),
\;\;\; \forall  \Lambda_+'(0) < \ell  < \Lambda_+'(\bar\theta),
\; \forall \ell' \neq \ell,
\end{eqnarray}
where $\theta(\ell)$ $\in$ $(0,\bar\theta)$ is s.t.
$\Lambda_+'(\theta(\ell))$ $=$ $\ell$ $\in$ $(\Lambda_+'(0),\Lambda_+'(\bar\theta))$.
Moreover, $\Lambda_+^*$ is continuous on $(-\infty,\Lambda_+'(\bar\theta))$.

\vspace{2mm}

\noindent {\it Step 2: Upper bound.}
For all $\ell$ $\in$ $\R$, $\pi$ $\in$ $\Ac$,
an application of Chebycheff's inequality yields:
\begin{eqnarray*}
\P[\bar L_T^\pi \geq \ell ] &\leq & \exp(-\theta \ell T) \E[\exp(\theta T \bar L_T^\pi)],
\;\;\; \forall \; \theta \in [0,\bar\theta),
\end{eqnarray*}
and so
\begin{eqnarray*}
\limsup_{T\rightarrow \infty} \frac{1}{T}
\ln \P[ \bar L_T^\pi \geq \ell ] &\leq & -\theta \ell +
\limsup_{T\rightarrow \infty} \frac{1}{T} \ln  \E[\exp(\theta T \bar L_T^\pi)],
\;\;\; \forall \; \theta \in [0,\bar\theta).
\end{eqnarray*}
By definitions of $\Lambda_+$ and  $\Lambda_+^*$, we deduce:
\begin{eqnarray} \label{upper}
\sup_{\pi\in\Ac} \limsup_{T\rightarrow \infty} \frac{1}{T}
\ln \P[\bar L_T^\pi \geq \ell  ] &\leq & - \Lambda_+^*(\ell).
\end{eqnarray}

\vspace{1mm}

\noindent {\it Step 3: Lower bound.}
Consider first the case  $\ell$ $\in$ $(\Lambda_+'(0),\Lambda_+'(\bar\theta))$, and 
let us define the probability measure $\Q_T$ on $(\Omega,\Fc_T)$  via:
\begin{eqnarray} \label{defQ}
\frac{d\Q_T}{d\P} &=& \exp\left[ \theta(\ell) L_T^{\pi^{+,\ell}} -
\Gamma_T(\theta(\ell),\pi^{+,\ell}) \right],
\end{eqnarray}
where 
\begin{eqnarray*}
\Gamma_T(\theta,\pi) &=&
\ln \E[\exp(\theta T \bar L_T^\pi)], \;\;\; \theta \in [0,\bar\theta), \;
\pi \in \Ac.
\end{eqnarray*}
For any $\eps$ $>$ $0$, we have: 
\begin{eqnarray*}
\frac{1}{T} \ln \P[ \ell - \eps < \bar L_T^{\pi^{+,\ell}} < \ell + \eps ] 
&=& \frac{1}{T} \ln\left(\int \frac{d\P}{d\Q_T}
1_{\left\{ \ell - \eps < \bar L_T^{\pi^{+,\ell}}
<  \ell +  \eps \right\}} d\Q_T \right) \\
&\geq& -\theta(\ell)\big(\ell + \eps \big)
+ \frac{1}{T} \Gamma_T(\theta(\ell),\pi^{+,\ell}) \\
& & \;\;\; + \frac{1}{T} \ln \Q_T \big[ \ell - \eps < \bar L_T^{\pi^{+,\ell}} < \ell +  \eps \big],
\end{eqnarray*}
where we use \reff{defQ} in the last inequality.
By definition of the dual problem, this yields:
\begin{eqnarray}
\liminf_{T\rightarrow\infty}
\frac{1}{T} \ln \P[\ell - \eps < \bar L_T^{\pi^{+,\ell}} < \ell + \eps   ] &\geq& -\theta(\ell)\big(\ell + \eps \big)
+ \Lambda_+(\theta(\ell))  \nonumber  \\
& & \;\;\; + \liminf_{T\rightarrow\infty}
\frac{1}{T} \ln \Q_T \left[ \ell - \eps < \bar L_T^{\pi^{+,\ell}} < \ell +  \eps  \right] \nonumber \\
&\geq& - \Lambda_+^*(\ell) - \theta(\ell) \eps  \nonumber \\
& &  + \liminf_{T\rightarrow\infty}
\frac{1}{T} \ln \Q_T \left[ \ell - \eps < \bar L_T^{\pi^{+,\ell}} < \ell +  \eps   \right],
\label{interlower1}
\end{eqnarray}
where the second inequality follows by the definition of $\Lambda_+^*$ (and actually holds with equality due to \reff{Lam*}).
We now show that:
\begin{eqnarray} \label{inter0}
\liminf_{T\rightarrow\infty}
\frac{1}{T} \ln \Q_T
\left[ \ell - \eps < \bar L_T^{\pi^{+,\ell}} < \ell +  \eps   \right] &=& 0.
\end{eqnarray}
Denote by $\tilde\Gamma_T$ the c.g.f.  under $\Q_T$ of $L_T^{\pi^{+,\ell}}$.  For all $\zeta$ $\in$ $\R$, we have by \reff{defQ}:
\begin{eqnarray*}
\tilde\Gamma_T(\zeta) &:=&
\ln \E^{\Q_T} [\exp(\zeta L_T^{\pi^{+,\ell}})] \\
&=& \Gamma_T(\theta(\ell) + \zeta,\pi^{+,\ell}) -
\Gamma_T(\theta(\ell),\pi^{+,\ell}).
\end{eqnarray*}
Therefore, by definition of the dual control problem  \reff{defLambda},
we have for all $\zeta$ $\in$ $[-\theta(\ell),\bar\theta - \theta(\ell))$:
\begin{eqnarray} \label{inter1}
\limsup_{T\rightarrow\infty} \frac{1}{T} \tilde\Gamma_T(\zeta) &\leq&
\Lambda_+(\theta(\ell)+\zeta) - \Lambda_+(\theta(\ell)).
\end{eqnarray}
As in part 1) of this proof, by Chebycheff's inequality, we have for
all $\zeta$ $\in$ $[0,\bar\theta-\theta(\ell))$:
\begin{eqnarray*}
\limsup_{T\rightarrow \infty} \frac{1}{T}
\ln \Q_T \left[\bar L_T^{\pi^{+,\ell}} \geq \ell + \eps \right]
&\leq & -\zeta( \ell +  \eps)
+ \limsup_{T\rightarrow \infty} \frac{1}{T} \tilde\Gamma_T(\zeta) \\
&\leq& -\zeta\left( \ell + \eps \right)
+ \Lambda_+(\zeta+\theta(\ell)) - \Lambda_+(\theta(\ell)),
\end{eqnarray*}
where the second inequality follows from \reff{inter1}. We deduce
\begin{eqnarray}
\limsup_{T\rightarrow \infty} \frac{1}{T}
\ln \Q_T \left[\bar L_T^{\pi^{+,\ell}} \geq \ell + \eps  \right]
&\leq& - \sup\{ \zeta \left( \ell + \eps \right)
- \Lambda_+(\zeta): \zeta \in [\theta(\ell),\bar\theta) \} \nonumber \\
& & \;\;\; - \Lambda_+(\theta(\ell))
+ \theta(\ell)\left( \ell + \eps \right) \nonumber \\
&\leq&  - \Lambda_+^*\left( \ell +  \eps \right)
- \Lambda_+(\theta(\ell)) + \theta(\ell)\left( \ell +  \eps \right), \nonumber \\
&=& - \Lambda_+^*\left( \ell + \eps \right) +
\Lambda_+^*(\ell) + \eps \theta(\ell), \label{inter2}
\end{eqnarray}
where the second inequality and the last equality follow
from \reff{Lam*}.
Similarly, we have for all $\zeta$ $\in$ $[-\theta(\ell),0]$:
\begin{eqnarray*}
\limsup_{T\rightarrow \infty} \frac{1}{T}
\ln \Q_T \left[\bar L_T^{\pi^{+,\ell}} \leq \ell - \eps \right]
&\leq & -\zeta \left( \ell - \eps \right)
+ \limsup_{T\rightarrow \infty} \frac{1}{T} \tilde\Gamma_T(\zeta) \\
&\leq& -\zeta \left( \ell - \eps \right)
 + \Lambda_+(\theta(\ell)+\zeta) - \Lambda_+(\theta(\ell)),
\end{eqnarray*}
and so:
\begin{eqnarray}
\limsup_{T\rightarrow \infty} \frac{1}{T}
\ln \Q_T \left[\bar L_T^{\pi^{+,\ell}} \leq \ell - \eps  \right]
&\leq& - \sup\{ \zeta \left( \ell - \eps \right)
- \Lambda_+(\zeta): \zeta \in [0,\theta(\ell)] \} \nonumber \\
& & \;\;\; - \Lambda_+(\theta(\ell))
+ \theta(\ell)\left( \ell - \eps \right) \nonumber \\
&\leq& - \Lambda_+^*\left( \ell - \eps \right)
+ \Lambda_+^*(\theta(\ell))  -  \eps \theta(\ell). \label{inter3}
\end{eqnarray}
By \reff{inter2}-\reff{inter3}, we then get:
\begin{eqnarray*}
& & \limsup_{T\rightarrow \infty} \frac{1}{T}
\ln \Q_T \left[ \left\{\bar L_T^{\pi^{+,\ell}}\leq \ell-\eps \right\}
\cup \left\{\bar L_T^{\pi^{+,\ell}}\geq  \ell + \eps  \right\}\right] \\
&\leq& \max \left\{
\limsup_{T\rightarrow \infty} \frac{1}{T}
\ln \Q_T \left[\bar L_T^{\pi^{+,\ell}} \geq \ell + \eps  \right];
\limsup_{T\rightarrow \infty} \frac{1}{T}
\ln \Q_T \left[\bar L_T^{\pi^{+,\ell}} \leq \ell - \eps \right]
\right\} \\
&\leq& \max \left\{
- \Lambda_+^*\left( \ell + \eps \right) +
\Lambda_+^*(\ell) + \eps \theta(\ell);
- \Lambda_+^*\left( \ell - \eps \right)
+ \Lambda_+^*(\theta(\ell))  -  \eps \theta(\ell) \right\} \\
& < & 0,
\end{eqnarray*}
where the strict inequality follows from \reff{resultexposed}.
This implies that $\Q_T[\{\bar L_T^{\pi^{+,\ell}}\leq \ell - \eps\}$
$\cup$ $\{\bar L_T^{\pi^{+,\ell}}\geq \ell +\eps\}]$
$\rightarrow$ $0$ and hence  $\Q_T [ \ell -\eps <\bar L_T^{\pi^{+,\ell}}< \ell + \eps ]$
$\rightarrow$ $1$ as $T$ goes to infinity. In particular \reff{inter0}
is satisfied, and by sending $\eps$ to zero in \reff{interlower1}, we
get for any $\ell'$ $<$ $\ell$ $<$  $\Lambda_+'(\bar\theta)$: 
\begin{eqnarray*}
\liminf_{T\rightarrow\infty}
\frac{1}{T} \ln \P[\bar L _T^{\pi^{+,\ell}} >  \ell' ] &\geq& \lim_{\eps\rightarrow 0}  \liminf_{T\rightarrow\infty} \frac{1}{T} \ln \P[ \ell - \eps < \bar L _T^{\pi^{+,\ell}} < \ell + \eps ] \\
& \geq &   - \Lambda_+^*(\ell).
\end{eqnarray*}
By continuity of $\Lambda_+^*$ on $(-\infty,\Lambda_+'(\bar \theta))$, we obtain  
\begin{eqnarray*}
\liminf_{T\rightarrow\infty} \frac{1}{T} \ln \P[\bar L _T^{\pi^{+,\ell}} \geq \ell ] &\geq&
- \Lambda_+^*(\ell).
\end{eqnarray*}
This last inequality combined with \reff{upper} proves the assertion for $v_+(\ell)$ when $\ell$ $\in$ $(\Lambda_+'(0),\Lambda_+'(\bar\theta))$. 

Now, consider the case $\ell$ $\leq$ $\Lambda_+'(0)$, and define $\ell_n$ $=$ $\Lambda_+'(0)+\frac{1}{n}$, $\pi^{+(n)}$ $=$ $\hat\pi(\theta(\ell_n))$. 
Then, by the same arguments as in \reff{interlower1} with $\ell_n$ $\in$ $(\Lambda_+'(0),\Lambda_+'(\bar\theta))$, we have
\beqs
\liminf_{T\rightarrow\infty} \frac{1}{T} \ln \P[\bar L _T^{\pi^{+(n)}} \geq \ell ]  & \geq & 
 \lim_{\eps\rightarrow 0}  \liminf_{T\rightarrow\infty} \frac{1}{T} \ln \P[ \ell_n - \eps < \bar L _T^{\pi^{+(n)}} < \ell_n + \eps ] \\ 
 & \geq &  -  \Lambda_+^*(\ell_n).
\enqs
By sending $n$ to infinity, together with the continuity of $\Lambda_+^*$, we get
\beqs
\liminf_{n\rightarrow\infty} \liminf_{T\rightarrow\infty} \frac{1}{T} \ln \P[\bar L _T^{\pi^{+(n)}} \geq \ell ] & \geq & - \Lambda_+^*(\Lambda_+'(0)) \; = \; 0,
\enqs
which  combined with \reff{upper}, ends the proof. 
\ep

\begin{Remark} \label{remdual+}
{\rm Theorem \ref{theodual+} shows that the upside chance large deviations control problem  can be solved via the resolution of the dual control problem. 
When the target growth rate level $\ell$ is  smaller than $\Lambda_+'(0)$, then one can achieve almost surely over long term an average growth term above $\ell$, in the sense that $v_+(\ell)$ $=$ $0$,  with a nearly optimal portfolio strategy which does not depend on this level. When the target level $\ell$  lies between 
$\Lambda_+'(0)$ and $\Lambda_+'(\bar\theta)$,  the optimal strategy depends on this level and is obtained from the optimal  strategy 
for the dual control problem $\Lambda_+(\theta)$ at point $\theta$ $=$ $\theta(\ell)$. When $\Lambda_+'(\bar\theta)$ $=$ $\infty$, i.e. $\Lambda_+$ is steep, we have a complete resolution 
of the large deviations control problem for all values of $\ell$. Otherwise,  the problem remains open  for $\ell$ $>$ $\Lambda_+'(\bar\theta)$. 
\ep
}
\end{Remark}

\vspace{3mm}

 Let us next consider the downside risk probability, and define the corresponding dual control problem:
 \beq \label{defLambda-}
 \Lambda_-(\theta) & := & \inf_{\pi\in\Ac}  \liminf_{T\rightarrow\infty} \frac{1}{T}  \ln \E \big[ e^{\theta T \bar L_T^\pi} \big], \;\;\; \theta \leq 0. 
 \enq
 Convexity of $\Lambda_-$ is not so straightforward as for $\Lambda_+$, and requires  the additional condition that the set of admissible controls  $\Ac$ is convex.  Indeed,  under this condition, we observe  from the dynamics \reff{Xpi} that a convex combination of wealth process is a wealth process. Thus, for any $\theta_1$, $\theta_2$ $\in$ $(-\infty,0)$, $\lambda$ $\in$ $(0,1)$, 
 $\pi^1$, $\pi^2$ $\in$ $\Ac$,  there exists $\pi$ $\in$ $\Ac$ such that:
 \beqs
 \frac{\lambda\theta_1}{\lambda\theta_1+(1-\lambda)\theta_2} X_T^{\pi^1}  +  \frac{(1-\lambda)\theta_2}{\lambda\theta_1+(1-\lambda)\theta_2} X_T^{\pi^2}  &=& X_T^\pi. 
 \enqs
 By concavity of the logarithm function, we then obtain
 \beqs
\ln X_T^\pi & \geq & \frac{\lambda\theta_1}{(\lambda\theta_1+(1-\lambda)\theta_2)}   \ln X_T^{\pi^1} +\frac{ (1-\lambda)\theta_2}{(\lambda\theta_1+(1-\lambda)\theta_2)}   \ln X_T^{\pi^2}, 
 \enqs
 and so, by setting $\theta$ $=$ $\lambda\theta_1+(1-\lambda)\theta_2$ $<$ $0$: 
 \beqs
\theta  T \bar  L_T^\pi & \leq & \lambda\theta_1 T \bar L_T^{\pi^1} +  (1-\lambda)\theta_2 T \bar  L_T^{\pi^2}.  
 \enqs
Taking exponential and expectation on both sides of this relation, and using H\"older inequality, we get:
\beqs
\E \big[ e^{\theta T \bar L_T^\pi} \big] & \leq & \Big(\E \big[  e^{\theta_1 T \bar L_T^{\pi^1}} \big] \Big)^\lambda \Big(\E \big[  e^{\theta_2 T \bar L_T^{\pi^2}} \big] \Big)^{1-\lambda}. 
\enqs
Taking logarithm, dividing by $T$, sending $T$ to infinity, and since $\pi^1$, $\pi^2$ are arbitrary in $\Ac$, we obtain by definition of $\Lambda_-$:
\beqs
\Lambda_-(\theta) & \leq & \lambda \Lambda_-(\theta_1) + (1-\lambda) \Lambda_-(\theta_2), 
\enqs
i.e. the convexity of $\Lambda_-$ on $\R_-$.  Since $\Lambda_-(0)$ $=$ $0$, the convex function $\Lambda_-$ is either infinite on $(-\infty,0)$ or 
finite on $\R_-$.  We now state the duality relation for downside risk large deviations probability, whose  proof  can be found in \cite{nag12}.

\begin{Theorem} \label{theodual-}
 Suppose that $\Lambda_-$ is  differentiable on $(-\infty,0)$, and there exists $\hat\pi(\theta)$ $\in$ $\Ac$ solution to 
 $\Lambda_-(\theta)$ for any $\theta$ $<$ $0$. Then, for all $\ell$ $<$ $\Lambda_-'(0)$, we have:
 \beqs
 v_-(\ell) &=& \inf_{\theta \leq 0} \big[ \Lambda_-(\theta) - \theta\ell \big], 
 \enqs
 and an optimal control for $v_-(\ell)$, when  $\ell$ $\in$ $(\Lambda_-'(-\infty),\Lambda_-'(0))$ is: 
 \beqs
 \pi^{-,\ell} &=&  \hat\pi(\theta(\ell)), \;\;\; \mbox{ with } \; \Lambda_-'(\theta(\ell)) \; = \; \ell,
 \enqs
 while $v_-(\ell)$ $=$ $-\infty$ when $\ell$ $<$ $\Lambda_-'(-\infty)$. 
 \end{Theorem}

\begin{Remark} \label{remdual-}
{\rm Theorem \ref{theodual-} shows that the downside risk large deviations control problem  can be solved via the resolution of the dual control problem.  
When the target growth rate level $\ell$ is  smaller than $\Lambda_-'(-\infty)$, then one can find a portfolio strategy so that   the average growth term almost never fall below $\ell$ over the long term, 
in the sense that  $v_-(\ell)$ $=$ $-\infty$.  When the target level $\ell$  lies between  $\Lambda_-'(-\infty)$ and $\Lambda_-'(0)$,  the optimal strategy depends on this level and is obtained from the optimal  strategy 
for the dual control problem $\Lambda_-(\theta)$ at point $\theta$ $=$ $\theta(\ell)$.  
\ep
}
\end{Remark}

\vspace{5mm}

\noindent {\bf Interpretation of the dual problem}

\noindent For $\theta$ $\neq$ $0$, the dual  problem  can be written as
\beqs
\frac{1}{\theta} \Lambda_\pm(\theta)  &=& \sup_{\pi \in \Ac} \limsup_{T\rightarrow\infty} J_T(\theta,\pi), 
\enqs
with
\beqs
J_T(\theta,\pi) & :=&  \frac{1}{\theta T} \ln \E \big[ e^{\theta T \bar L_T^\pi} \big],  
\enqs
and is known in the literature as a risk-sensitive control problem.  A Taylor expansion around $\theta$ $=$ $0$  puts in evidence the role played by the 
risk sensitivity parameter $\theta$:
\beqs
J_T(\theta,\pi)  & \simeq  & \E\big[ \bar L_T^\pi \big] + \theta T {\rm Var}(\bar L_T^\pi) + O(\theta^2).  
\enqs
This relation shows that risk-sensitive control amounts to making dynamic the Markowitz  problem: one maximizes the expected average growth rate subject to a constraint on its variance. 
Risk-sensitive portfolio criterion on finite horizon $T$ has been studied  in  \cite{biepli99} and \cite{davleo08}, and in the ergodic case $T$ $\rightarrow$ $\infty$, by 
\cite{fleshe00} and \cite{nagpen02}.

 \vspace{5mm}

\noindent {\bf  Endogenous utility function}

\noindent  Recalling that growth rate is the logarithm of wealth process, the duality relation   for the upside large deviations probability 
means formally that for large horizon $T$: 
\beqs
\P \big[ \bar L_T^{\pi^+,\ell} \geq \ell \big] & \simeq & \exp\big( v_+(\ell) T \big)  \\
&=& \exp \big( \Lambda_+(\theta(\ell)) T - \theta(\ell)\ell T \big) \\
& \simeq & \E \Big[ \big( X_T^{\pi^{+,\ell}} \big)^{\theta(\ell)} \Big] e^{-\theta(\ell) T}, \;\;\; \mbox{ with } \; \theta(\ell) > 0.
\enqs
Similarly, we have for the downside risk probability: 
\beqs
\P \big[ \bar L_T^{\pi^{-,\ell}} \leq \ell \big] & \simeq & \E \Big[ \big( X_T^{\pi^{-,\ell}} \big)^{\theta(\ell)} \Big] e^{-\theta(\ell) T}, \;\;\; \mbox{ with } \; \theta(\ell) < 0.
\enqs 
In other words, the target growth rate level $\ell$ determines endogenously  the risk aversion parameter $1-\theta(\ell)$ of an agent with Constant Relative Risk Aversion (CRRA) utility function and large investment horizon.  Moreover, the optimal strategy $\pi^{\pm,\ell}$ for $v_\pm(\ell)$ is expected to provide a good approximation for the solution to the CRRA utility maximization problem 
\beqs
\sup_{\pi\in\Ac} \E \big[ (X_T^\pi)^{\theta(\ell)} \big], 
\enqs
with a large but finite time horizon.

\section{A toy model: the Black Scholes case}

\setcounter{equation}{0} \setcounter{Assumption}{0}
\setcounter{Theorem}{0} \setcounter{Proposition}{0}
\setcounter{Corollary}{0} \setcounter{Lemma}{0}
\setcounter{Definition}{0} \setcounter{Remark}{0}

We illustrate the results of the previous section in a toy example, namely the Black-Scholes model,  with one stock of price process
\beqs
dS_t &=& S_t \big( b d + \sigma dW_t \big), \;\;\; t \geq 0. 
\enqs
We also consider an agent with constant proportion portfolio  strategies. In other words, the set of admissible controls $\Ac$ is equal to $\R$. 
Given a constant proportion $\pi$ $\in$ $\R$ invested in the stock, and starting w.l.o.g. with unit capital, the average growth rate portfolio of the agent  is equal to
\beqs
\bar L_T^\pi \; = \; \frac{L_T^\pi}{T} &=& \big( b \pi - \frac{\sigma^2\pi^2}{2} \big) + \sigma \pi \frac{W_T}{T}. 
\enqs
It follows that $\bar L_T^\pi$ is distributed according to a Gaussian law: 
\beqs
\bar L_T^\pi & \leadsto & \Nc\Big( b\pi - \frac{\sigma^2\pi^2}{2} \; , \frac{\sigma^2\pi^2}{T}  \Big),
\enqs
and its (limiting) Log-Laplace function is equal to
\beqs
\Gamma(\theta,\pi) & := & (\lim_{T\rightarrow\infty}) \frac{1}{T} \ln \E \Big[ e^{\theta T \bar L_T^\pi} \Big]  
\; = \;  \theta \Big[  b \pi - (1-\theta) \frac{\sigma^2\pi^2}{2}  \Big]
\enqs

$\bullet$ {\bf Upside chance probability}. 

\noindent The dual control problem in the upside case is then given by 
\beqs
\Lambda_+(\theta) \; = \;  \sup_{\pi\in\R}  \Gamma(\theta,\pi) &=& 
\left\{ 
\begin{array}{cl}
\infty, & \mbox{ if } \theta\geq 1, \\
\Gamma(\theta,\hat\pi(\theta)) \; = \; \frac{b^2}{2\sigma^2} \frac{\theta}{1-\theta}, & \mbox{ if } 0 \leq \theta < 1,
\end{array}
\right.
\enqs
with 
\beqs
\hat\pi(\theta) &=& \frac{b}{\sigma^2(1-\theta)}. 
\enqs
Hence, $\Lambda_+$ differentiable on $[0,1)$ with: $\Lambda_+'(0)$ $=$ $\frac{b^2}{2\sigma^2}$, and 
$\Lambda_+'(1)$ $=$ $\infty$, i.e. $\Lambda_+$ is steep.   From Theorem \ref{theodual+}, the value function of the upside large deviations probability is 
explicitly computed as:
\beqs
v_+(\ell) &:=& \sup_{\pi\in\R}   \limsup_{T\rightarrow\infty} \frac{1}{T} \ln \P\big[ \bar L_T^\pi \geq \ell  \big] \\
&=& \inf_{0\leq\theta < 1} \big[ \Lambda_+(\theta) - \theta\ell \big] \\
&=& \left\{
\begin{array}{cl}
0, & \mbox{ if } \; \ell  \leq \Lambda_+'(0) \; = \;  \frac{b^2}{2\sigma^2}  \\
- \big( \sqrt{\Lambda_+'(0)} - \sqrt{\ell} \big)^2, &   \mbox{ if } \; \ell   >  \Lambda_+'(0)
\end{array}
\right.
\enqs
with an optimal strategy:
\beqs
\pi^{+,\ell} &=&  \left\{
\begin{array}{cl}
\frac{b}{\sigma^2} , & \mbox{ if } \; \ell  \leq  \Lambda_+'(0)   \\
& \\
\sqrt{\frac{2\ell}{\sigma^2}}, &   \mbox{ if } \; \ell   >   \Lambda_+'(0). 
\end{array}
\right. 
\enqs
Notice that, when $\ell$ $\leq$ $\Lambda_+'(0)$, we have not only a nearly optimal control as stated in Theorem \ref{theodual+}, but an optimal control 
given by $\pi^+$ $=$ $b/\sigma^2$, which is precisely the optimal portfolio  for the classical Merton problem with logarithm utility function. 
Indeed, in this model, we have by the law of large numbers: 
$\bar L_T^{\pi^+}$ $\rightarrow$ $\frac{b^2}{2\sigma^2}$ $=$ $\Lambda_+'(0)$, as $T$ goes to infinity, and so 
$\lim_{T\rightarrow\infty} \frac{1}{T} \ln \P[\bar L_T^{\pi^+} \geq \ell ]$ $=$ $0$ $=$ $v_+(\ell)$. Otherwise, when $\ell$ $>$ $\Lambda_+'(0)$, the optimal strategy depends on $\ell$, and the larger  the target growth rate level, the more one has to invest in the stock.

\vspace{2mm}

$\bullet$ {\bf Downside risk  probability}. 

\noindent  The dual control problem in the downside case is then given by 
\beqs
\Lambda_-(\theta) \; = \;  \inf_{\pi\in\R}  \Gamma(\theta,\pi) &=& 
 \Gamma(\theta,\hat\pi(\theta)) \; = \; \frac{b^2}{2\sigma^2} \frac{\theta}{1-\theta}, \;\;\; \theta \leq 0, 
\enqs
with 
\beqs
\hat\pi(\theta) &=& \frac{b}{\sigma^2(1-\theta)}. 
\enqs
Hence, $\Lambda_-$ is differentiable on $\R_-$  with: $\Lambda_-'(-\infty)$ $=$ $0$, and  $\Lambda_-'(0)$ $=$ $\frac{b^2}{2\sigma^2}$.  
From Theorem \ref{theodual+}, the value function of the downside large deviations probability is 
explicitly computed as:
\beqs
v_-(\ell) &:=& \inf_{\pi\in\R}   \liminf_{T\rightarrow\infty} \frac{1}{T} \ln \P\big[ \bar L_T^\pi \leq \ell  \big] \\
&=& \inf_{\theta \leq 0} \big[ \Lambda_-(\theta) - \theta\ell \big] \\
&=& \left\{
\begin{array}{cl}
- \infty, & \mbox{ if } \; \ell < 0 \\
- \big( \sqrt{\Lambda_-'(0)} - \sqrt{\ell} \big)^2, &   \mbox{ if } \; 0 \leq \ell   \leq  \Lambda_-'(0) \; = \;  \frac{b^2}{2\sigma^2} 
\end{array}
\right.
\enqs
with an optimal strategy:
\beqs
\pi^{-,\ell} &=&  
 \sqrt{\frac{2\ell}{\sigma^2}},   \;  \mbox{ if } \; 0 \leq \ell   \leq  \Lambda_-'(0). 
\enqs
Moreover,  when $\ell$ $<$ $0$, and by choosing $\pi^-$ $=$ $0$, we have $\bar L_T^{\pi^-}$ $=$ $0$, so that $\P[\bar L_T^{\pi^-} \leq \ell]$ $=$ $0$, and thus $v_-(\ell)$ $=$ $-\infty$. In other words, 
when the target growth rate $\ell$ $<$ $0$, by doing nothing, we have an optimal strategy for $v_-(\ell)$.

\vspace{3mm}

\begin{Remark}
{\rm  The above direct calculations  rely on the fact that we restrict portfolio $\pi$ to be constant in proportion.  Actually, the explicit forms of the value function and optimal strategy remain the same if  we allow  a priori  portfolio strategies $\pi$ $\in$ $\Ac$  to change over time based on the available information, i.e to be $\F$-predictable.  This requires more advanced tools from stochastic control and PDEs to be presented  in the sequel   in a more general framework. 
\ep
}
\end{Remark}

\section{Factor model}

\setcounter{equation}{0} \setcounter{Assumption}{0}
\setcounter{Theorem}{0} \setcounter{Proposition}{0}
\setcounter{Corollary}{0} \setcounter{Lemma}{0}
\setcounter{Definition}{0} \setcounter{Remark}{0}

We consider a market model with one riskless asset price $S^0$ $=$ $1$, and $d$ stocks of price process $S$ governed by
\beqs
dS_t &=& {\rm diag}(S_t)\big( b(Y_t) dt + \sigma(Y_t) dW_t) \\
dY_t &=& \eta(Y_t) dt + \gamma(Y_t) dW_t,
\enqs
where $Y$ is a factor process valued in $\R^m$, and $W$ is a $d+m$ dimensional standard Brownian motion.  The coefficients $b,\sigma,\eta,\gamma$ are assumed to satisfy regular conditions ensuring existence of a unique strong solution to the above stochastic differential equation, and $\sigma$ is also of full rank, i.e. the $d\times d$-matrix $\sigma\sigma'$ is invertible.

A portfolio strategy $\pi$ is an $\R^d$-valued adapted process, representing the fraction of wealth invested in the $d$ stocks.  The admissibility condition for $\pi$ in $\Ac$ will be precised later, but for the moment 
$\pi$ is required to satisfy the integrability conditions: 
\beqs
\int_0^T |\pi_t'b(Y_t)| dt + \int_0^T |\pi_t'\sigma(Y_t)|^2 dt &<& \infty,  \;\; a.s.  \mbox{ for all } \; T > 0.
\enqs
The growth rate portfolio 
is then given by: 
\beqs
L_T^\pi &=& \int_0^T \Big( \pi_t'b(Y_t) - \frac{\pi_t'\sigma\sigma'(Y_t)\pi_t}{2} \Big) dt  + \int_0^T \pi_t'\sigma(Y_t) dW_t. 
\enqs
For any $\theta$ $\in$ $\R$, and $\pi$, we  compute the Log-Laplace function of the growth rate portfolio:
\beqs
\Gamma_T(\theta,\pi) & :=& \ln \E \big[ e^{\theta L_T^\pi} \big]  \\
&=& \ln \E \Big[ \Ec\Big(\int_0^T \theta \pi_t'\sigma(Y_t)dW_t\Big) e^{\theta \int_0^T f(\theta,Y_t,\pi_t) dt } \Big],
\enqs
where $\Ec(.)$ denotes the  Dol\'eans-Dade exponential, and $f$ is the function:
\beqs
f(\theta,y,\pi) &=&  \pi'b(y) - \frac{1-\theta}{2} \pi'\sigma\sigma'(y)\pi. 
\enqs
We now impose the admissibility condition that $\pi$ lies in $\Ac$ if the Dol\'eans-Dade local martingale $\Ec\Big(\int_0^. \theta \pi_t'\sigma(Y_t)dW_t\Big)_{0\leq t\leq T}$ is a true martingale for any $T$ $>$ $0$, 
which is ensured, for instance,  by the Novikov condition.  In this case, this Dol\'eans-Dade exponential defines a probability measure $\Q_\pi$ equivalent to $\P$ on $(\Omega,\Fc_T)$,  and  we have:
\beqs
\Gamma_T(\theta,\pi) &=& \ln \E^{\Q_\pi} \Big[ \exp{\big(\theta \int_0^T f(\theta,Y_t,\pi_t) dt } \big)\Big],
\enqs
where  $Y$ is governed under $\Q_\pi$ by
\beqs
dY_t &=& \big( \eta(Y_t) + \theta \gamma(Y_t)\sigma'(Y_t)\pi_t \big) dt + \gamma(Y_t) dW_t^\pi. 
\enqs
with $W^\pi$ a $\Q_\pi$-Brownian motion from Girsanov's theorem. 

\vspace{1mm}

We then consider the dual control problems:

\begin{itemize}
\item {\bf Upside chance}: for $\theta$ $\geq$ $0$,
\beqs
\Lambda_+(\theta) &=& \sup_{\pi \in \Ac} \limsup_{T\rightarrow\infty} \frac{1}{T} \ln \E^{\Q_\pi} \Big[ \exp{\big(\theta\int_0^T f(\theta,Y_t,\pi_t) dt\big) } \Big].
\enqs
\item {\bf Downside risk}: for $\theta$ $\leq$ $0$, 
\beqs
\Lambda_-(\theta) &=& \inf_{\pi \in \Ac} \liminf_{T\rightarrow\infty} \frac{1}{T} \ln \E^{\Q_\pi} \Big[  \exp{\big(\theta\int_0^T f(\theta,Y_t,\pi_t) dt\big) }\Big]. 
\enqs
\end{itemize}
These problems are known in the literature as ergodic risk-sensitive  control problems, and studied by dynamic programming methods in \cite{benfre92},  \cite{flemac95} and \cite{nag96}. 
Let us now formally derive the ergodic equations associated to these risk-sensitive control problems.  We consider the finite horizon risk-sensitive stochastic control problems:
\beqs
u_+(T,y;\theta) &=&  \sup_{\pi\in\Ac} \E^{\Q_\pi} \Big[ \exp{\big(\theta\int_0^T f(\theta,Y_t,\pi_t) dt\big) } \big| Y_0 = y  \Big], \; \theta\geq 0 \\
u_-(T,y;\theta) &=&  \inf_{\pi\in\Ac} \E^{\Q_\pi} \Big[ \exp{\big(\theta\int_0^T f(\theta,Y_t,\pi_t) dt\big) } \big| Y_0 = y  \Big], \; \theta \leq 0, 
\enqs
and by using the formal substitution: 
\beqs
\ln u_\pm(T,y;\theta) & \simeq & \Lambda_\pm(\theta) T + \varphi_\pm(y;\theta), \;\;\; \mbox{ for large } T, 
\enqs
in the corresponding Hamilton-Jacobi-Bellman (HJB) equations for $u_\pm$:
\beqs
\DT{u_\pm} &=& \sup_{\pi\in\R^d} \big[ \theta f(\theta,y,\pi) u_\pm + (\eta(y)+\theta\gamma(y)\sigma'(y)\pi)'D_y u_\pm + 
\frac{1}{2}{\rm tr}(\gamma\gamma'(y)D_y^2 u_\pm) \big],
\enqs
we obtain the ergodic HJB equation for the pair $(\Lambda_\pm(\theta),\varphi_\pm(.,\theta))$ as: 
\beqs
\Gamma(\theta) &=& \eta(y)'D_y\varphi_{} + \frac{1}{2}{\rm tr}(\gamma\gamma'(y)D_y^2\varphi_{}) 
+ \frac{1}{2} \big| \gamma'(y)D_y \varphi_{} \big|^2 \\
& &  + \; \theta \sup_{\pi\in\R^d} \Big[ \pi'(b(y)+\sigma(y)\gamma'(y)D_y \varphi_{}\big) - \frac{1-\theta}{2}  \pi'\sigma\sigma'(y)\pi \Big], 
\enqs
which is well-defined for $\theta$ $<$ $1$.  In the above equation $\Gamma(\theta)$ is a candidate for $\Lambda_\pm(\theta)$ while $\varphi$ is a candidate solution for $\varphi_\pm$.  This can be rewritten as a semi-linear ergodic PDE with quadratic growth  in the gradient: 
\beq 
\Gamma(\theta) &=&  \big( \eta(y) + \frac{\theta}{1-\theta} \gamma\sigma'(\sigma\sigma')^{-1} b(y) \big).D_y\varphi  + \frac{1}{2}{\rm tr}(\gamma\gamma'(y)D_y^2\varphi_{}) \nonumber \\
& & \; + \frac{1}{2} D_y\varphi'\gamma(y)\big[ I_{_{d+m}} + \frac{\theta}{1-\theta} \sigma'(\sigma\sigma')^{-1}\sigma(y)\big]\gamma'(y)D_y\varphi \nonumber \\
& & \; + \ \frac{\theta}{2(1-\theta)} b'(\sigma\sigma')^{-1}b(y), \label{ergodicPDE}
\enq
and a candidate for optimal feedback  control of the dual problem: 
\beq \label{hatpi}
\hat\pi(y;\theta) &=& \frac{1}{1-\theta} (\sigma\sigma')^{-1}(y)\big[  b(y)+ \sigma\gamma'(y)D_y\varphi(y;\theta)\big].  
\enq
We now face  the questions:
\begin{itemize}
\item  Existence of a pair solution $(\Gamma(\theta),\varphi(.,\theta))$ to the ergodic  PDE \reff{ergodicPDE}?
\item  Do we have $\Gamma(\theta)$ $=$ $\Lambda_\pm(\theta)$, and what is the domain of $\Gamma$?
\end{itemize}

We give some  assumptions, which allows us to answer  the above issues.

\vspace{2mm}

{\bf (H1)}  \hspace{5mm}  $b$, $\sigma$, $\eta$ and $\gamma$ are smooth $C^2$ and globally Lipschitz. 

\vspace{2mm}

{\bf (H2)}  \hspace{5mm}  $\sigma\sigma'(y)$ and $\gamma\gamma'(y)$ are uniformly elliptic: there exist $\delta_1$, $\delta_2$ $>$ $0$ s.t.
\beqs
\delta_1 |\xi|^2 & \leq \; \xi' \sigma\sigma'(y) \xi \; \leq & \delta_2 |\xi|^2, \;\;\; \forall \xi, y \in \R^m, \\
\delta_1 |\xi|^2 & \leq \; \xi' \gamma\gamma'(y) \xi \; \leq & \delta_2 |\xi|^2, \;\;\; \forall \xi, y \in \R^m.  
\enqs

\vspace{2mm}

{\bf (H3)} \hspace{5mm}   There exist $c_1$ $>$ $0$ and $c_2$ $\geq$ $0$ s.t. 
\beqs
b(\sigma\sigma')^{-1}b(y) &\geq c_{1}|y|^2 - c_{2}, \;\;\; \forall y \in \R^m.
\enqs

\vspace{2mm}

{\bf (H4)} \hspace{3mm} Stability condition: there exist $c_3$ $>$ $0$ and $c_4$ $\geq$ $0$ s.t. 
\beqs
\big( \eta(y) - \gamma \sigma'(\sigma\sigma')^{-1}b(y) \big).y & \leq & - c_{3} |y|^2 +  c_{4}
\enqs

\vspace{2mm}

According to \cite{benfre92} (see also \cite{nag12} and \cite{robxin13}),  the next result states the existence of a smooth solution to the ergodic equation.

\begin{Proposition}
Under {\bf (H1)}-{\bf (H4)}, there exists for any $\theta$ $<$ $1$, a solution  $(\Gamma(\theta),\varphi(.;\theta))$ with $\varphi(.;\theta)$ $C^2$,  to the ergodic HJB equation s.t: 
\begin{itemize}
\item For $\theta$ $<$ $0$, $\varphi(.;\theta)$  is upper-bounded
\beqs
\varphi(y;\theta) & \longrightarrow & - \infty, \;\;\; \mbox{ as } \; |y| \rightarrow \infty,
\enqs
\item  For $\theta$ $\in$ $(0,1)$, $\varphi(.;\theta)$  is lower-bounded
\beqs
\varphi(y;\theta) & \longrightarrow &  \infty, \;\;\; \mbox{ as } \; |y| \rightarrow \infty,
\enqs
\end{itemize}
and 
\beqs
\big| D_y\varphi(y;\theta) \big| & \leq & C_\theta (1 + |y|). 
\enqs
\end{Proposition}

We now relate a solution to the ergodic equation to the dual risk-sensitive control problem. In other words, this means the convergence of the finite horizon risk-sensitive stochastic control 
to the component $\Gamma$ of the ergodic equation.  We distinguish the downside and upside cases.  

\vspace{2mm}

$\bullet$ {\bf Downside risk}:  In this case, it is shown in \cite{nag12} that for all $\theta$ $<$ $0$,  the solution $(\Gamma(\theta),\varphi(.;\theta)$ to \reff{ergodicPDE}, with 
$\varphi(.,\theta)$ $C^2$ and upper bounded, is unique (up to an additive constant for $\varphi(.;\theta)$), and we have:
\beqs
\Gamma(\theta) &=& \Lambda_-(\theta), \;\;\; \theta < 0. 
\enqs
Moreover, there is an admissible optimal feedback control $\hat\pi(.,\theta)$  for $\Lambda_-(\theta)$ given by \reff{hatpi}, and for which the factor process $Y$ is ergodic under $\Q_{\hat\pi}$. 
It is also proved in \cite{nag12} that $\Gamma$ $=$ $\Lambda_-$ is differentiable on $(-\infty,0)$.  Therefore, from Theorem \ref{theodual-},  the solution to the 
downside risk large deviations probability is given by: 
\beqs
v_-(\ell) &=&  \inf_{\theta \leq 0} \big[ \Gamma(\theta) - \theta\ell \big], \;\;\; \ell < \Gamma'(0), 
\enqs
with an optimal control:
\beqs
\pi_t^{-,\ell} &=& \hat \pi(Y_t;\theta(\ell)), \;\;\; \Gamma'(\theta(\ell)) \; = \; \ell, \;\;\; \forall \ell \in (\Gamma'(-\infty),\Gamma'(0)),
\enqs
while $v_-(\ell)$ $=$ $-\infty$ for $\ell$ $<$ $\Gamma'(-\infty)$. 

\vspace{2mm}

$\bullet$ {\bf  Upside chance}:   In this case, $0<\theta<1$, there is no unique  solution $(\Gamma(\theta),\varphi(.;\theta))$ to the ergodic equation, with $\varphi(.;\theta)$ $C^2$ lower-bounded, even up 
to an additive constant, as pointed out in \cite{fleshe00}.  In general, we only have a verification type result, which states that if the process $Y$ is ergodic under $\Q_{\hat\pi}$, then 
\beqs
\Gamma(\theta) &=& \Lambda_+(\theta),
\enqs
and $\hat\pi(.,\theta)$ is an optimal feedback control for $\Lambda_+(\theta)$.  

\vspace{3mm}

In the next paragraph, we consider a  linear factor model for which explicit calculations can be derived.

\subsection{Linear Gaussian factor model}

We consider the linear factor model:
\beqs
dS_t &=& {\rm diag}(S_t) \big( (B_1Y_t + B_0) dt + \sigma dW_t) \;\;\; \mbox{ in } \;  \R^d, \\
dY_t &=& KY_t dt + \gamma dW_t, \;\;\; \mbox{ in } \;  \R^m,
\enqs 
with $K$ a stable matrix in $\R^m$,  $B_1$ a constant $d\times m$ matrix, $B_0$ a non-zero vector in $\R^d$, $\sigma$ a $d\times (d+m)$-matrix of rank $d$, and $\gamma$ a nonzero $m\times (d+m)$ matrix.  
We are searching for a candidate solution to the ergodic equation \reff{ergodicPDE} in the quadratic form: 
\beqs
\varphi(y;\theta) &=& \frac{1}{2} C(\theta)y.y + D(\theta)y, \;\;\; y \in \R^m, 
\enqs
for some  $m\times m$ matrices  $C(\theta)$ and $D(\theta)$.  Plugging this form of $\varphi$ into \reff{ergodicPDE}, we find that $C(\theta)$ must solve the algebraic Riccati equation:
\beq
 \frac{1}{2}C(\theta)'\gamma\big(I_{d+m} + \frac{\theta}{1-\theta} \sigma'(\sigma\sigma')^{-1}\sigma\big)\gamma' C(\theta)  & & \nonumber \\
 +  \; \big( K+  \frac{\theta}{1-\theta} \gamma\sigma'(\sigma\sigma')^{-1}B_1\big)' C(\theta) +  \frac{1}{2} \frac{\theta}{1-\theta} B_1'(\sigma\sigma')^{-1}B_1 &=& 0, \label{riccati}
\enq
while $B(\theta)$ is determined by
\beqs
\Big( K + \frac{\theta}{1-\theta} \gamma\sigma'(\sigma\sigma')^{-1}B_1 + \gamma \big(I_{d+m} + \frac{\theta}{1-\theta} \sigma'(\sigma\sigma')^{-1}\sigma\big)\gamma' C(\theta)  \Big)' D(\theta) & & \\
 + \;  \frac{\theta}{1-\theta} \big( \sigma \gamma' C(\theta) + B_1)' (\sigma\sigma')^{-1} B_0 &=& 0.
\enqs
Then, $\Gamma(\theta)$ is  given by:
\beqs
\Gamma(\theta) &=& \frac{1}{2} {\rm tr}(\gamma\gamma'C(\theta))  + \frac{1}{2} D(\theta)'\gamma(I_{d+m}+ \frac{\theta}{1-\theta}\sigma'(\sigma\sigma')^{-1}\sigma)\gamma'D(\theta) \\
& & \; + \; \frac{\theta}{1-\theta}  B_0' (\sigma\sigma')^{-1} \sigma \gamma' D(\theta)  + \frac{1}{2} \frac{\theta}{1-\theta}  B_0'(\sigma\sigma')^{-1}B_0,
\enqs
and a candidate for the optimal feedback control is: 
\beqs
\hat\pi(y;\theta) &=& \frac{1}{1-\theta} (\sigma\sigma')^{-1}\big[  (B_1+ \sigma\gamma' C(\theta))y  + B_0 + \sigma\gamma' D(\theta) \big].  
\enqs
In \cite{fleshe00}, it is shown that there exists some positive $\bar\theta$ small enough, s.t.  for $\theta$ $<$ $\bar\theta$,  
there exists a solution $C(\theta)$ to the Riccati equation \reff{riccati} s.t. $Y$ is ergodic under $\Q_{\hat\pi}$, and so by verification theorem, 
$\Gamma(\theta)$ $=$ $\Lambda_\pm(\theta)$.  In the one-dimensional  asset and factor model, as studied in \cite{pha03}, we obtain more precise results. Indeed, in this case: $d$ $=$ $m$ $=$ $1$, the Riccati equation is a second-order polynomial equation in $C(\theta)$, which admits two  explicits roots given by:
\beqs
C_\pm(\theta) &=& - \frac{K}{|\gamma|^2} \left[ \frac{  1 - \theta\big(1  -  \rho \frac{|\gamma| B_1}{K|\sigma|}\big)  \pm  \sqrt{ (1-\theta)(1-\theta \beta)}} { 1 -\theta(1-\rho^2) } \right],
\enqs 
for all $\theta$ $\leq$ $\bar\theta$, with 
\beqs
\bar\theta \; = \; \frac{1}{\beta} \wedge 1, & &  \beta \; = \; 1 - \rho^2 + \Big( \rho -   \frac{ |\gamma| B_1}{ K |\sigma|  }\Big)^2 \; > \; 0, 
\enqs
where $|\gamma|$ (resp. $|\sigma|$) is the Euclidian norm of $\gamma$ (resp. $\sigma$), and $\rho$ $\in$ $[-1,1]$ is the correlation between $S$ and $Y$, i.e.  $\rho$ $=$ $\frac{\gamma\sigma'}{|\gamma||\sigma|}$. 
Actually, only the solution  $C(\theta)$ $=$ $C_-(\theta)$ is relevant in the sense that for this  root, $Y$ is ergodic under $Q_{\hat\pi}$, and  thus  by verification theorem: 
\beqs
\Lambda_\pm(\theta) \; = \;  \Gamma(\theta) & = & \frac{1}{2}|\gamma|^2 C_-(\theta)  + \frac{1}{2} |\gamma|^2 D(\theta)^2 \big( 1 +  \frac{\theta}{1-\theta}\rho^2 \big) \\
& & \; + \;  \frac{\theta}{1-\theta} \frac{B_0}{|\sigma|} \rho |\gamma| D(\theta)  + \frac{1}{2} \frac{\theta}{1-\theta} \frac{B_0^2}{|\sigma|^2}, \;\;\; \theta < \bar\theta,
\enqs
where 
\beqs
D(\theta) &=& - \frac{B_0}{K|\sigma|} \frac{  \theta \big(\rho |\gamma| C_-(\theta) + \frac{B_1}{|\sigma|} \big)} {\sqrt{(1-\theta)(1-\theta\beta)}},
\enqs
and with optimal control for $\Lambda_\pm(\theta)$ given by:
\beqs
\hat\pi(y;\theta) &=& \frac{1}{(1-\theta)|\sigma|} \Big[ \Big( \frac{B_1}{|\sigma|} + \rho|\gamma| C_-(\theta) \Big) y + \frac{B_0}{|\sigma|} + \rho |\gamma| D(\theta) \Big].  
\enqs
Moreover, it is also proved in \cite{pha03},  that 
\beqs
\Gamma'(0) &=& \frac{B_0^2}{2|\sigma|^2} -  \frac{B_1^2 |\gamma|}{4 |\sigma|^2 K} \; > \; 0,
\enqs
(recall that $K$ $<$ $0$) and  the function $\Gamma$ is steep, i.e. 
\beqs
\lim_{\theta\uparrow \bar\theta} \Gamma'(\theta) &=& \infty. 
\enqs
From Theorems \ref{theodual+} and \ref{theodual-},  the solutions to the upside chance and downside risk large deviations probability are given by: 
\beqs
v_+(\ell) &=&  \inf_{0  \leq  \theta < \bar\theta} \big[ \Gamma(\theta) - \theta\ell \big], \;\;\; \ell \in \R, \\
v_-(\ell) &=&  \inf_{\theta \leq 0} \big[ \Gamma(\theta) - \theta\ell \big], \;\;\; \ell < \Gamma'(0), 
\enqs
with  optimal control and nearly optimal control  for $v_+(\ell)$:
\beqs
\pi_t^{+,\ell} &=& \hat \pi(Y_t;\theta(\ell)), \;\;\; \Gamma'(\theta(\ell)) \; = \; \ell, \;\;\;  \mbox{ when }  \ell > \Gamma'(0), \\
\pi_t^{+(n)} &=& \hat\pi(Y_t;\theta_n), \;\;\;  \mbox{ with } \; \theta_n = \theta(\Gamma'(0) + \frac{1}{n}) \overset{n\rightarrow\infty}{\longrightarrow} 0, \;\;\;  \mbox{ when } \ell \leq \Gamma'(0), 
\enqs
and optimal control for $v_-(\ell)$:
\beqs
\pi_t^{-,\ell} &=& \hat \pi(Y_t;\theta(\ell)), \;\;\; \Gamma'(\theta(\ell)) \; = \; \ell, \;\;\; \forall \ell \in (\Gamma'(-\infty),\Gamma'(0)).  
\enqs

\subsection{Examples}

\noindent $\bullet$ {\bf Black-Scholes model.} This corresponds to the case where $B_1$ $=$ $0$. Then, $\beta$ $=$ $\bar\theta$ $=$ $1$, $C_-(\theta)$ $=$ $D(\theta)$ $=$ $0$, and so
\beqs
\Lambda_\pm(\theta) \; = \; \Gamma(\theta) &=&  \frac{1}{2} \frac{\theta}{1-\theta} \frac{B_0^2}{|\sigma|^2}, \;\;\; \forall \theta < 1. 
\enqs  
We thus obtain the same optimal strategy as described in Section 4. 
 
 \vspace{3mm}
 
 \noindent $\bullet$ {\bf Platen-Rebolledo  model.} In this model, the logarithm of the stock price $S$  is governed by an Ornstein-Uhlenbeck process $Y$, and this corresponds to the case where 
 $B_1$ $=$ $K$ $<$ $0$,  $B_0$ $=$ $\frac{1}{2}|\gamma|^2$ $>$ $0$, $\gamma$ $=$ $\sigma$, and thus $\rho$ $=$ $1$. Then, $\beta$ $=$ $0$, $\bar\theta$ $=$ $1$, 
 \beqs
 C_-(\theta) \; = \;  \frac{|K|}{|\sigma|^2}\big[ 1 - \sqrt{1-\theta} \big], & & D(\theta) \; = \; -  \frac{1}{2}  \theta,
 \enqs
 and so
 \beqs
 \Gamma(\theta) &=&  \frac{|K|}{2} \big[ 1 - \sqrt{1-\theta} \big] + \theta \frac{|\sigma|^2}{8},  \;\;\; \theta < 1, \\
 \Gamma'(0) & = &  \bar \ell \; := \;  \frac{|K|}{4} + \frac{|\sigma|^2}{8}, \;\;\;\;\; \Gamma'(-\infty) \; = \;  \underline\ell \; := \; \frac{|\sigma|^2}{8}, 
 \enqs
 \beqs
 \theta(\ell) &=& 1 - \Big( \frac{\bar\ell - \underline\ell}{\ell -\underline\ell} \Big)^2, \;\;\; \forall \ell > \underline\ell. 
 \enqs
The solution to the upside chance large deviations probability is then given by:
\beqs
v_+(\ell) & =&    \left\{ 
			\begin{array}{cl} 
			- \frac{(\ell-\bar\ell)^2}{\ell - \bar\ell + \frac{|K|}{4}}, & \mbox{ if } \; \ell  > \bar\ell  \\
			0, & \mbox{ if } \; \ell \leq \bar\ell.
			\end{array}
			\right. 
\enqs
with optimal (resp. nearly optimal)  portfolio strategy:
\beqs
\pi_t^{+,\ell} &=&    \frac{ K- 4(\ell-\bar\ell)}{|\sigma|^2} Y_t +  \frac{1}{2}, \;\;\;  \mbox{ if } \;  \ell  > \bar\ell  \\
\pi_t^{+(n)} &=&  \frac{K - 1/n}{|\sigma|^2} Y_t +  \frac{1}{2}, \;\;\;  \mbox{ if } \; \ell  \leq \bar\ell.
\enqs 
The solution to the downside risk  large deviations probability is  given by:
\beqs
v_-(\ell) & =&    \left\{ 
			\begin{array}{cl} 
			- \frac{(\ell-\bar\ell)^2}{\ell - \underline\ell}, & \mbox{ if } \;  \underline \ell < \ell  \leq  \bar\ell  \\
			- \infty, & \mbox{ if } \; \ell \leq \underline\ell,
			\end{array}
			\right. 
\enqs
with optimal  portolio strategy:
\beqs
\pi_t^{-,\ell} &=&   - \frac{4(\ell-\underline\ell)}{|\sigma|^2} Y_t +  \frac{1}{2}, \;\;\;  \mbox{ if } \; \underline\ell < \ell \leq  \bar\ell  
\enqs

\end{document}